\documentclass[twocolumn]{aastex631}
\usepackage{amsmath}
\usepackage{units}
\usepackage{listings}
\usepackage{wasysym}
\usepackage{natbib}
\usepackage{bm}
\usepackage{colortbl}
\definecolor{LightCyan}{rgb}{0.88,1,1}

\NewDocumentCommand{\codeword}{v}{%
\texttt{\textcolor{blue}{#1}}%
}

\begin{document}

\title{Predicting the rate of fast radio bursts in globular clusters from binary black hole observations}

\author[0009-0002-5780-4539]{Aryamann Rao}
\affiliation{Department of Astronomy and Astrophysics, University of Toronto, 50 St George St, Toronto, ON M5S 3H4, Canada}
\correspondingauthor{Aryamann Rao}
\email{aryamann.rao@gmail.com}

\author[0000-0001-9582-881X]{Claire S. Ye}
\affiliation{Canadian Institute for Theoretical Astrophysics, University of Toronto, 60 St George Street, Toronto, ON M5S 3H8, Canada}

\author[0000-0002-1980-5293]{Maya Fishbach}
\affiliation{Canadian Institute for Theoretical Astrophysics, University of Toronto, 60 St George Street, Toronto, ON M5S 3H8, Canada}

\begin{abstract} 
The repeating fast radio burst (FRB) source in an old globular cluster (GC) in M81 proves that FRBs, which are typically associated with young magnetars, can also occur in old stellar populations. A potential explanation is super-Chandrasekhar binary white dwarf (BWD) coalescences, which may produce FRB-emitting neutron stars. GCs can also give rise to binary black hole (BBH) mergers detectable with gravitational waves, and the BWD coalescence rate from GCs is correlated with their BBH merger rate. For the first time, we combine independent observations of gravitational waves and FRBs to infer the origins of FRB sources. We use GC formation histories inferred from BBH observations to predict the rate of super-Chandrasekhar BWD coalescences originating from GCs as a function of redshift. We explore mass-loss and mass-conserved scenarios for BWD coalescences and find that the coalescence rates evolve differently across redshift in these two cases. In the mass-loss scenario, the BWD coalescence rates decrease with increasing redshift, similar to some recent measurements of the FRB rate as a function of redshift. We show that GCs could contribute $\lesssim 1\%$ to the total FRB source formation rates in the local Universe. Our multi-messenger approach also offers a novel method to better constrain the GC population using both FRB and gravitational wave observations.
\end{abstract}

\section{Introduction} \label{sec:intro}

Fast radio bursts (FRBs) are bright, millisecond-long radio flares that have been observed to originate from extra-Galactic sources, with some sources giving rise to repeating bursts \citep{Lorimer_2007,Spitler_2014,Marcote_2017}. In recent years, the discovery rate of these mysterious signals has accelerated. For instance, the Parkes radio telescope is credited to numerous FRB detections including the first detection (namely the Lorimer burst) in 2007 \citep{Lorimer_2007,Keane_2012,Thornton_2013,Burke_2014,Petroff_2015, Ravi_2015,Champion_2016}. Detections have also been made by the Arecibo telescope \citep{Spitler_2014} and the Green Bank telescope \citep{gbt_2015}. The Canadian Hydrogen Intensity and Mapping Experiment (CHIME) Telescope discovered and catalogued more than 500 FRBs in 2021 \citep{CHIME_Amiri_2021}. The Australian Square Kilometre Array Pathfinder (ASKAP) recently reported the detection of a very distant FRB at redshift $z>1$ \citep{askap_2023}. Despite the numerous detections, there is no common consensus amongst the community on the sources of these bursts.

The redshift evolution of FRB rates may provide strong indications of their origins. On one hand, some studies argue that the rate of FRBs closely follows the star formation history \citep[SFH; e.g.,][]{James2022,Bhardwaj+2024_hostgalaxy,Wang_vanLeeuwen_2024}, implying a connection to short-lived, massive stars. This is supported by the recent detection of FRB200428 from the Galactic soft gamma repeater SGR 1935+2154 \citep{galactic_magnetar,CHIME_GalacticFRB}, which clearly shows that some FRBs are generated by violent activities on the surface of young magnetars (\citealp{Thompson_Duncan_1995,Kashiyama+2013,Popov_Postnov_2013}; \citealp[see][for a review]{Petroff_2022}). Magnetars are young neutron stars (NS) with extremely strong magnetic fields ($\sim 10^{14}$ G), traditionally thought to form by the collapse of massive stars in core-collapse supernovae. Magnetars radiate their energy over a span of a few thousand years, which causes their magnetic fields to decay until they reach regular NS levels \citep[see][for a review of magnetars]{Kaspi_2017}. Hence, if magnetars formed via core-collapse supernovae contribute to the majority of the observed FRBs, the FRB rate as a function of redshift must trace the SFH of the Universe, which increases with increasing redshift in the local Universe before peaking at $z \approx 2$.

On the other hand, other studies find that the FRB rate as a function of redshift has a significant departure from the SFH and hence cannot be entirely attributed to young stellar remnants \citep[e.g.,][]{Hashimoto+2022,Qiang+2022,Zhang_Zhang_2022,Zhang2024,Law+2024}. This is supported by the discovery of a repeating FRB source detected in a globular cluster (GC) in the galaxy M81 \citep{Bhardwaj_2021,Kirsten_2022}. GCs are very old clusters of stars with no ongoing star formation (for a review on GCs, see \citealt{Gratton_2019}). These observations can be explained by postulating that either magnetars are not the only source of FRBs or that young NSs within GCs form via other channels, such as the coalescences of massive binary white dwarfs (BWD; \citealp[e.g.,][]{Schwab_2016,Schwab_2021,Lu+2022,Kremer+2021_frb}). In either case, the FRB rate would be significantly delayed with respect to the SFH.

In this study, we investigate how much the latter case, i.e. young NSs produced by BWD coalescences in dense star clusters, contribute to the observed FRB rate. We compare the formation rate of young NSs from BWD coalescences in realistic GC simulations to the inferred rate of FRBs both as a function of redshift. We obtain the latter from previous studies, most of which are based on modeling CHIME FRB observations. The predicted BWD coalescence rates from GCs depend on GC population properties throughout cosmic history, including their formation rate and distributions in mass and size, which are highly uncertain. These GC formation histories also affect the population of binary black hole (BBH) mergers. Thus, gravitational wave (GW) observations of BBH mergers can probe the uncertain GC population, with consequences for the FRB rate from GCs. In this work, we use GW constraints on the GC population from~\citet{Fishbach_Fragione_2023} to predict the BWD coalescence rate from GCs and compare to the FRB rate as a function of redshift. Our results build on \citet{Kremer_2023_2}, who also predicted the BWD coalescence rate as a function of redshift using realistic GC simulations. In our study, we improve the predictions for the redshift evolution of the BWD coalescence rate by using the GC population properties inferred from GW observations. We compare our findings to a wide range of recent measurements of the FRB rate versus redshift.

This paper is structured as follows. In \autoref{sec:gc populations} we describe the realistic simulations of GCs and the delay time distributions of BBH and BWD coalescences in these simulations (\autoref{subsec:gc simulations}). We also discuss the parameters governing GC formation histories and property distributions (\autoref{subsec:gc formation history}). \autoref{sec:bwd rate and frb rate} discusses the BWD coalescence rates inferred from the aforementioned factors in \autoref{sec:gc populations} (\autoref{subsec:derive bwd rate}). Here we also describe the various FRB rate models from the literature that we will consider for this study (\autoref{subsec:frb rate}) and compare our inferred BWD coalescence rates to the FRB rate models (\autoref{subsec:compare bwd and frb rate}). \autoref{sec:discussion} presents a discussion of the sources of uncertainty in our results and \autoref{sec:conclusion} lists the key takeaways of our work.

\section{Population of Globular Clusters}\label{sec:gc populations}
We utilize a large grid of realistic simulations of GCs run with the state-of-the-art Cluster Monte Carlo (\texttt{\texttt{CMC}}) $N$-body code \citep{Kremer_2020} to study the BWD coalescence rates from GCs. We assume the GC population follows the constraints derived from GW observations~\citep[][their Figure~B1 and B2]{Fishbach_Fragione_2023}. These constraints apply to a set of parameters that define the GC population properties, which are further described in \autoref{subsec:gc formation history} below.

\subsection{Simulations of globular clusters}\label{subsec:gc simulations}

\begin{figure*}
    \centering
    \includegraphics[width=\textwidth]{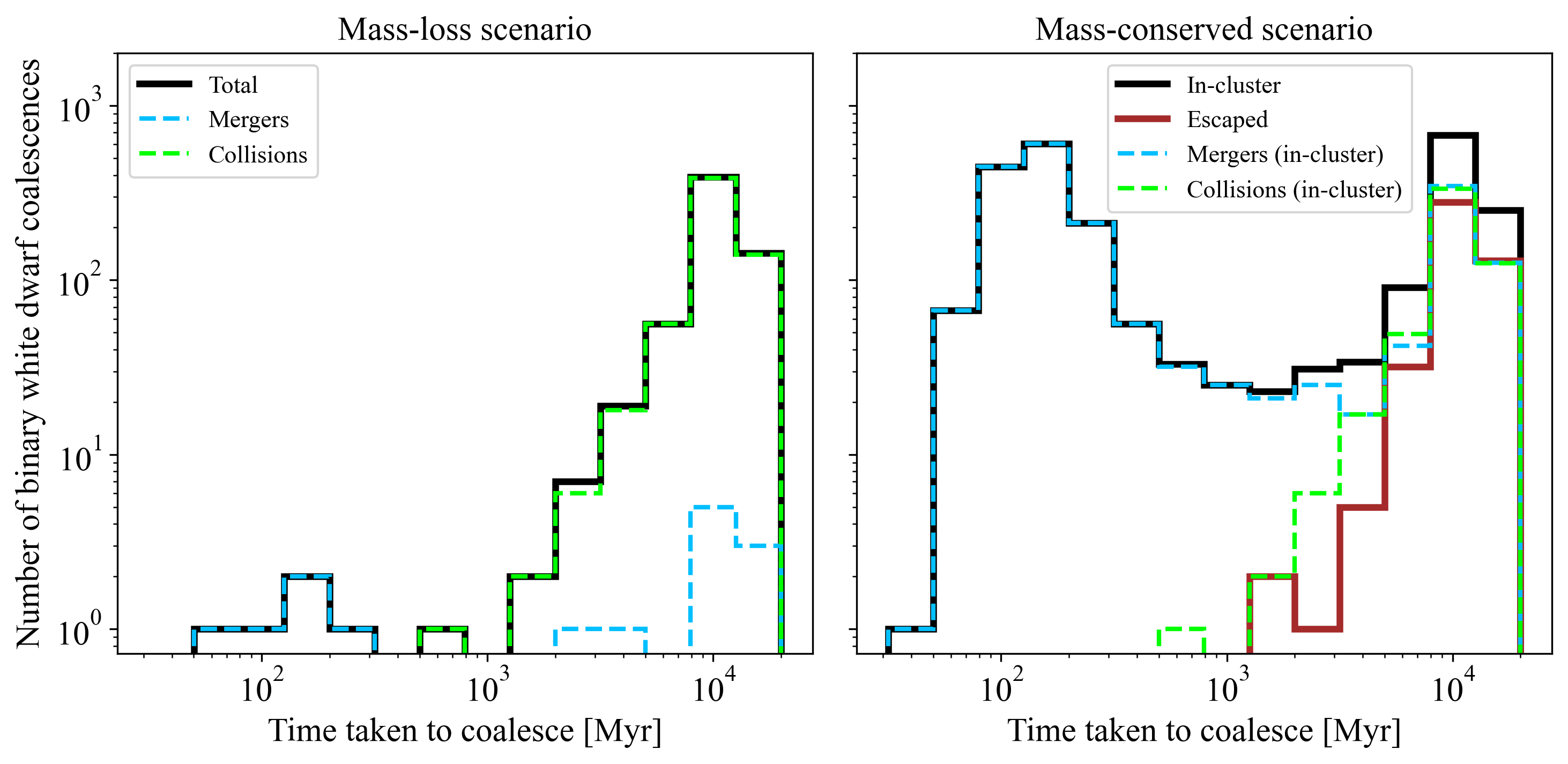}
    \caption{
    The histograms show the delay time distributions of the super-Chandrasekhar BWD coalescences that collapse to NSs. \emph{Left:} The distribution of delay times of all BWD to NS coalescences in the `mass-loss' scenario (black histogram). The resulting distribution when only mergers are considered is shown in dashed blue. When only collisions are considered, the distribution is shown in dashed green. \emph{Right:} The distribution of delay times for all BWD to NS coalescences that occured within the cluster in the `mass-conserved' scenario is shown in black. The resulting distribution when only in-cluster mergers are considered is shown in dashed blue. When only in-cluster collisions are considered, the distribution is shown in dashed green. Furthermore, the resulting distribution when only escaped coalescences (BWD that merge outside the clusters) are considered is shown in solid brown. Hence the blue and green distributions add up to the black distribution in each sub-figure.
    } 
    \label{fig:bwd delay time dist}
\end{figure*}

Each \texttt{CMC} simulation is defined by four initial cluster properties: 

\begin{itemize}
    \item Mass (with $M\in\{1.2, 2.4, 4.8, 9.6\}\times 10^5\ M_{\astrosun}$)
    \item Metallicity (with $Z\in\{0.01, 0.1, 1\}\ Z_{\astrosun}$)
    \item Virial Radius (with $r_v\in \{0.5, 1, 2, 4\}\ \text{pc}$)
    \item Galactocentric distance (with $R_g\in~\{2, 8, 20\}\ \text{kpc}$)
\end{itemize}

This gives a total of $144$ simulations with every combination of the above-mentioned properties. Tracked in each simulation are the binary and dynamical interactions of the stellar constituents. The stellar initial mass function is assumed to follow the Kroupa standard broken power law \citep{Kroupa_2001} between $0.08\ M_{\astrosun}$ and $150\ M_{\astrosun}$. The initial binary fraction for all stars is set to $5\%$ and the mass of the secondary star is drawn from a flat distribution in mass ratio in the range $[0.1, 1]$ \citep[e.g.,][]{Duquennoy_Mayor_1991}.

For this study, we are interested in all the collisions and mergers between two white dwarfs (WDs) that can result in FRB-emitting young NSs \citep[e.g.,][]{Lu+2022,Kremer+2023} in \texttt{CMC}. `Collision' here refers to two initially unbound objects directly colliding with each other during dynamical encounters (the pericenter distance of the encounter is smaller than the sum of the WD radii), whereas `merger' refers to the eventual inspiral and merging of gravitationally bound binary systems. We assume that all super-Chandrasekhar collisions and mergers between a pair of Carbon-Oxygen (CO) WDs and/or Oxygen-Neon (ONe) WDs collapse to NSs \citep[e.g.,][]{Kremer2021,Ye+2024}.  For the rest of the paper, we use the umbrella term of `coalescence' to refer to both types of events unless a distinction needs to be made.

\begin{figure*}
    \centering
    \includegraphics[width=0.9\textwidth]{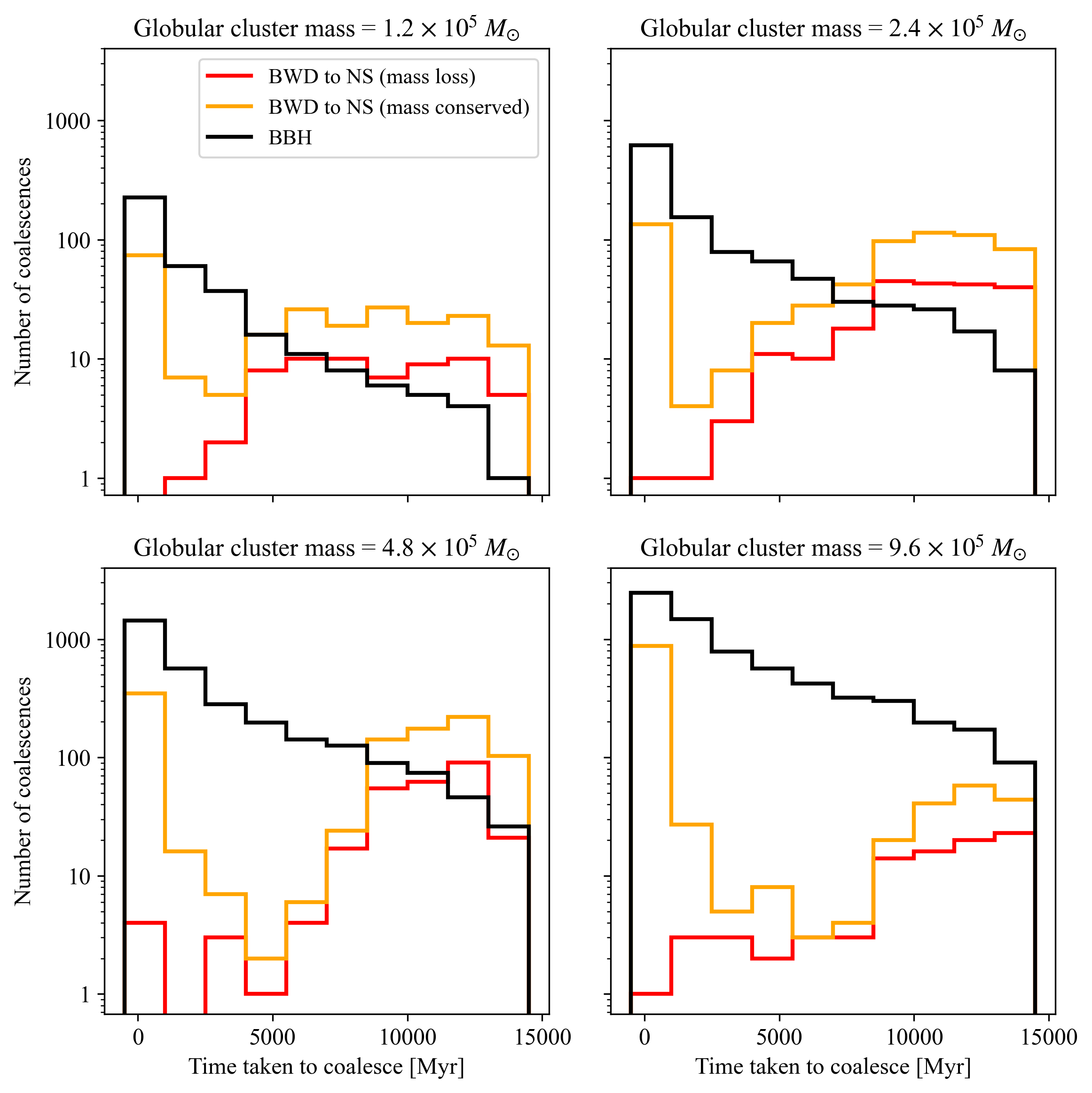}
    \caption{Each sub-figure plots the delay time distributions of three types of events originating from all 36 GCs for a given initial GC mass in the \texttt{CMC} simulations, (summing over the different virial radii, Galactocentric distances, and metallicities with equal weights). The events considered are i) BWD coalescences that result in NSs according to the mass-loss scenario, ii) BWD coalescences that result in NSs according to the mass-conserved scenario, and iii) BBH coalescences.
    }
    \label{fig:bwd_and_bbh_dist}
\end{figure*}

We consider two scenarios for dynamically unstable mass transfer during BWD mergers that can be used for determining the final remnants. As a lower-limit scenario, we adopt the default prescription of mass transfer in \citet{cosmic}. According to this prescription, the more massive WD (say of mass $M_1$) accretes total mass of $\Delta M$, which for typical $1 M_{\odot}$ WDs is approximately $10^{-3}\ M_{\odot}$. The less massive WD is destroyed in this merger. If $M_1 + \Delta M> 1.4\ M_{\odot}$ then the remnant can collapse to a NS from merger-induced collapse. In other cases the remnant is a CO/ONe WD. (for a detailed summary of the physics of BWD mergers/collisions in the simulations see \citealp{cosmic}; \citealp{Kremer_2020}; \citealp{Kremer2021}). 

As an alternative upper-limit scenario, we assume that mass is conserved during BWD mergers following, e.g., \citet[][for a more detailed discussions about this assumption, see \citealp{Ye+2024}]{Schwab_2021}. Hence a merger results in a NS if the total mass of the progenitors is greater than the Chandrasekhar mass ($1.4\ M_{\astrosun}$). This means we also include cases where the lower-limit scenario classifies the final remnant as a WD but the upper-limit scenario classifies it as a NS. For brevity, we will refer to the former lower-limit scenario as the `mass-loss' scenario and the latter upper-limit scenario as the `mass-conserved' scenario. We perform our analysis and report our results on these two limits separately. 

For the BWD collisions, we adopt a `sticky sphere' approximation as in \citet{Kremer_2020}, leading to collision products with total masses equal to the sum of the component masses. Hence the number of collisions are the same in both the aforementioned scenarios.

The distributions of the delay times of the BWD coalescences resulting in NSs are shown in \autoref{fig:bwd delay time dist}. The delay time refers to the time elapsed since GC formation. It can be seen that for the mass-conserved scenario, there are a lot more events with short delay times compared to the mass-loss scenario (see the peak closer to the delay time of $100$~Myr in the right panel of \autoref{fig:bwd delay time dist}, which is not present in the left panel). This is because the mass-conserved scenario allows more BWD mergers from primordial binaries to collapse to NSs at early times of the clusters' evolution, while a few BWD mergers can produce NSs in the mass-loss scenario \citep[also see][their Figure~7]{Ye+2024}.\footnote{Because of the negligible number of NSs formed from BWD mergers in the mass-loss scenario, we do not take into account the BWDs that merge outside the clusters in this scenario.} In general, BWD mergers (both from primordial binaries and dynamically-formed binaries) have delay times ranging from about $10^2$ to $10^4$~Myr. In comparison, collisions (always dynamically mediated) mostly show peaks at long delay times of about $10^4$~Myr. Since we are only concerned with BWD coalescences that result in NSs, from here on the term `BWD coalescence' refers specifically to those coalescences that form NSs.

Critically for our work, the rates and delay times of BWD coalescences are related to the population of BBH coalescences originating from GCs. \autoref{fig:bwd_and_bbh_dist} considers each initial GC mass used in the \texttt{CMC} simulations and plots the delay time distributions of BWD coalescences (in both mass-loss and mass-conserved scenarios) along with BBH coalescences. For a given GC mass, we sum over the number of coalescences in the 36 \texttt{CMC} simulations at that mass with equal weights.

In each subpanel, we see the following general trend. At short delay times (the earliest stages after GC formation), there are approximately 500 BBH coalescences for every mass-loss BWD coalescence. As the GC ages (longer delay times), we get more mass-loss BWD coalescences, in most cases exceeding the number of BBH coalescences with long delay times. For GCs with initial masses of $1.2\times 10^5\ M_{\odot}$ and $2.4\times 10^5\ M_{\odot}$, the ratio of BBH to mass-loss BWD coalescences is close to $1:5$ for long delay times. For GCs with initial mass $4.8\times 10^5\ M_{\odot}$, this ratio is closer to $1:1$, and for GCs of initial mass $9.6\times 10^5\ M_{\odot}$, the ratio is about $5:1$, where the number of BBH coalescences exceed the number of BWD coalescences even at long delay times. 

The trends in \autoref{fig:bwd_and_bbh_dist} can be attributed to mass segregation in GCs, which leads to more massive objects like black holes (BHs) sinking to the GC centre earlier in a GC's life. This increases the chances of dynamical encounters between BHs, resulting in a lot more BBH coalescences at early stages (short delay times). 
As part of these dynamical encounters, BHs from the cores get ejected out of the clusters in a phenomenon termed `BH burning' \citep{bh_burning}. Eventually, less massive objects like WDs sink towards the core of the GCs, allowing them to have increased dynamical encounters at late times (longer delay times for coalescences).

In addition to the dynamically-assembled BWD coalescences, there is a significant number of mass-conserved BWD mergers at short delay times. The ratio of BBH to BWD coalescences in the mass-conserved scenario is close to $2:1$ for short delay times as compared to $500:1$ for the mass-loss scenario. This is because the mass-conserved scenarios allows a lot more BWDs from primordial binaries to form NSs as explained earlier. At longer delay times however, both scenarios show similar ratios of BBH to BWD coalescences.

Hence, the \texttt{CMC} simulations show a general trend that most BHs dynamically interact and merge before the WDs in GCs. This explains how GW measurements of the BBH coalescence rate can be used to calibrate the predictions for BWD coalescences from GCs. This is done by first constraining the GC formation histories using GW observations of merging BBHs. Because the GC formation histories influences the BWD coalescences, we can then obtain the BWD coalescence rate as a function of redshift. 

\newpage
\subsection{Globular cluster formation histories}\label{subsec:gc formation history}
To estimate the coalescence rates of a pair of WDs from GCs as a function of redshift, we utilize the initial distributions of cluster properties and the cluster formation histories estimated from GW detections of BBHs as in \citet{Fishbach_Fragione_2023}. Here we briefly describe each distribution and the parameters governing them. The posteriors on these parameters obtained by \citet{Fishbach_Fragione_2023} will be used in this study as outlined in later sections. 

\textit{Distribution of GC masses:} The initial mass distributions of GCs are assumed to follow the Schechter distribution \citep{Schechter1974} which is

\begin{equation}
    p(M)\propto \Big(\dfrac{M}{M_*}\Big)^{\beta_m}\exp\Big(-\dfrac{M}{M_*}\Big)\,,
    \label{eqn:schechter dist}
\end{equation}

\noindent where the variable parameters are $M_*$ and $\beta_m$. The Schechter distribution is assumed to cover GC masses over the adopted \texttt{CMC} mass range from $1.2\times10^5\,M_\odot$ to $9.6\times10^5\,M_\odot$.

\textit{Distribution of GC metallicities:}
We assume the GC metallicities follow a lognormal distribution where the mean of the distribution is taken to be redshift dependent. The distribution can be expressed as
\begin{equation}
    p(Z|z) \propto \exp\Big(-\dfrac{1}{2}\Big(\dfrac{\log Z - \log \mu(z)}{\sigma_Z}\Big)^2\Big)\,.
    \label{eqn:lognormal dist}
\end{equation}
The dependence of the log mean on the redshift is determined by the relation in \citet{Mandau_Fragos_2017},

\begin{equation}
    \log \mu(z) = 0.153 - 0.074z^{1.34}\,.
    \label{eqn:log mean}
\end{equation}

We take the metallicity variance to be $\sigma_Z = 0.5$ dex. Here it should be noted that the metallicity values while calculating the distribution in \autoref{eqn:lognormal dist} are in units of solar metallicity, $Z_{\astrosun}=0.02$. We do not vary any parameters in this distribution.

\textit{Distribution of GC virial radii:} The initial virial radius distribution is taken to be a Gaussian with mean $\mu_r$ and standard deviation $\sigma_r$. In other words,

\begin{equation}
    p(r_v) \propto \exp\Big(-\dfrac{1}{2}\Big(\dfrac{r_v - \mu_r}{\sigma_r}\Big)^2\Big)\,.
    \label{eqn:virial gaussian}
\end{equation}

The variable parameters in this case are the mean $\mu_r$ and standard deviation $\sigma_r$.

\textit{Distribution of GC galactrocentric distances:} We assume that all the galactrocentric distance values considered by \citet{Kremer_2020} are equally probable and hence take each probability as $\nicefrac{1}{3}$.

Along with the distributions of initial cluster properties, we also take into account the functional form of the GC formation history. \citet{Fishbach_Fragione_2023} constrained the posteriors on variables governing the rate of GC formation as a function of redshift using GW detections of BBH coalescences. This is mathematically expressed as \citep{Madau_Dickinson_2014},

\begin{equation}
    \mathcal{R}_{\text{GC}}(z) = \mathcal{R}_0\dfrac{(1+z)^{a_z}}{1+(\frac{1+z}{1+z_{\text{peak}}})^{a_z+b_z}}\,,
    \label{eqn:gc formation rate}
\end{equation}

\noindent where $a_z, b_z, z_{\text{peak}}$ are the free parameters. To determine the normalising constant $R_0$ we first integrate \autoref{eqn:gc formation rate} to a lookback time $t_{\text{max}}$ corresponding to a redshift of $20$. This gives us
\begin{equation}
    n_0 = \int_0^{t_{\text{max}}}R_{\text{GC}}(t)\ dt
    \label{eqn:integral gc formation}\,.
\end{equation}
The quantity $n_0$ equals the number density of GCs formed over time $t_{\text{max}}$ given a formation history of $R_{\text{GC}}(t)$. After formation, many GCs will dissolve through stellar evolution, two-body relaxation, and/or the tidal stripping of stars by the potential of their host galaxies. Hence the number density of surviving GCs $n_{\text{surv}}$ follows $n_{\text{surv}} < n_0$ and is assumed to be $2.31\times 10^9\ \text{Gpc}^{-3}$\citep[][and references therein]{Rodriguez+2016}. The parameter $f_{\text{ev}}= \nicefrac{n_0}{n_{\text{surv}}}$ is taken to be a free parameter. Given a value of $f_{\text{ev}}$ we can calculate $n_0$ which allows us to calculate $R_0$ based on \autoref{eqn:integral gc formation}. 

\citet{Fishbach_Fragione_2023} estimated the posterior distributions of the parameters $\{M_{*},\beta_m, \mu_r, \sigma_r, a_z, b_z, z_{\text{peak}}, f_{\text{ev}}\}$ (see Figures~B1 and B2 in \citealp{Fishbach_Fragione_2023}). For brevity, we will refer to this set as $S_{\text{param}}$. These distributions were calculated by fitting the BBH coalescences predicted by the \texttt{CMC} simulations from \citet{Kremer_2020} to GW observations. As a reference, we define a default parameter set where $M_*=~10^{6.26}\ M_{\astrosun}$, $\beta_m=~-2$, $\mu_r=~2\ \text{pc}$, $\sigma_r =~2\ \text{pc}$, $a_z=~2.6$, $b_z=~3.6$, $z_{\text{peak}}=~2.2$, and $f_{\text{ev}}=~7.3$. These values correspond to a BBH coalescence rate of $10\ \text{Gpc}^{-3}\ \text{yr}^{-1}$ at $z=0$ \citep[also see][]{Ye_and_Fishbach_2024}.

Here, since the parameter $f_{\text{ev}}$ in \citet{Fishbach_Fragione_2023} was estimated for GC masses ranging from $10^4$ to $10^8$ $M_{\odot}$, and the \texttt{CMC} simulations do not cover this whole mass range (see \autoref{subsec:gc simulations}), we apply a factor of 0.16 to $f_{\text{ev}}$ to account for the difference in the mass range. Additionally, revisiting the calculation by \citet{Fishbach_Fragione_2023}, we note that their $f_{\text{ev}}$ is smaller by a factor of 3 because they did not properly weight the three different galactocentric radii. In summary, the $f_{\text{ev}}$ we use in this work is 0.48 times the $f_{\text{ev}}$ values presented in \citet{Fishbach_Fragione_2023}.

\cite{Fishbach_Fragione_2023} present two sets of posteriors, one which assumes the mass distribution parameters ($M_{*}, \beta_m$) to be constant and the other which assumes the virial radius distribution parameters ($\mu_r, \sigma_r$) to be constant. When not fit for, the parameters take the default values mentioned above. Although we performed our analysis using both cases, we found that the final results do not show any noticeable difference in either case. Hence we shall only present our results obtained using the posteriors which take the mass distribution parameters ($M_{*}, \beta_m$) to be constant. 

To avoid confusion henceforth, we will refer to the variables such as GC mass, virial radius, metallicity, and galactrocentric distance as GC `properties' and the set of variables $S_{\text{param}}$ defining the distributions of these properties as GC `parameters'.
When calculating the BWD coalescence rates as a function of redshift, we marginalize over the GC population parameters, averaging over 500 $S_{\text{param}}$ posterior samples.

\section{Rate of Binary White Dwarf Coalescences and Fast Radio Bursts}\label{sec:bwd rate and frb rate}
\subsection{Predicting the rate of binary white dwarf coalescences}\label{subsec:derive bwd rate}

For any GC $i$ with a given set of parameters $\{M^i, r^i_{v}, Z^i, R_g^i\}$ we calculate its contribution to the rate of coalescence events $l$ ($l$ could be BWD coalescences, BBH coalescences, WD--NS coalescences and so on) as a function of redshift with the equation \citep{Fishbach_Fragione_2023},
\begin{multline}\label{eqn:event rate}
    \mathcal{R_{\text{event}}}(z) = P(R_g^i)P(M^i)P(r^i_v)\\
    \times \sum_{l}\mathcal{R_{\text{GC}}}(\hat{z}_l(t+\tau_l))P(Z^i|\hat{z}_l(t+\tau_l))\,.
\end{multline}
Here, $P$ denotes the probability of finding a cluster with the given properties (to distinguish it from the probability density $p$). 
For any property $A$, the probability is taken to be
\begin{equation}\label{eqn:probability computation}
    P(A^i) = \dfrac{p(A^i)}{\sum_j p(A^j)}\,,
\end{equation}
\noindent where the sum is over all possible values $A^j$. The probability of each galactrocentric distance value considered is taken to be $\nicefrac{1}{3}$, as mentioned earlier.
Meanwhile, $\hat{z}_l(t+\tau_l)$ denotes the redshift corresponding to the formation time $t+\tau_l$ of the clusters where $t$ is the lookback time for redshift $z$ (where we are performing our calculation) and $\tau_l$ is the delay time of the event $l$. To convert between the times and redshift, we use the Planck 2018 cosmology provided in the \codeword{astropy} package \citep{Planck_2018}.

Our approach was verified on the BBH coalescences data from the \texttt{CMC} simulations. For the default values of $S_{\text{param}}$ mentioned in \autoref{subsec:gc formation history}, we obtain a BBH coalescence rate curve that peaks at $z=1.4$. When normalised to $10\ \text{Gpc}^{-3}\ \text{yr}^{-1}$ at $z=0$, the peak rate equals $20\ \text{Gpc}^{-3}\text{yr}^{-1}$. This is in close agreement with the results obtained by \cite{Ye_and_Fishbach_2024} on the same BBH coalescence data (the black curve in Figure~2 of their paper). Running similar tests with GC parameter values used by \citet{Fishbach_Fragione_2023} we were able to re-obtain all the BBH coalescence rates given in Figure~2 of their paper.

\subsection{Rate of fast radio bursts}\label{subsec:frb rate}
\begin{figure}
    \centering
    \includegraphics[width=\columnwidth]{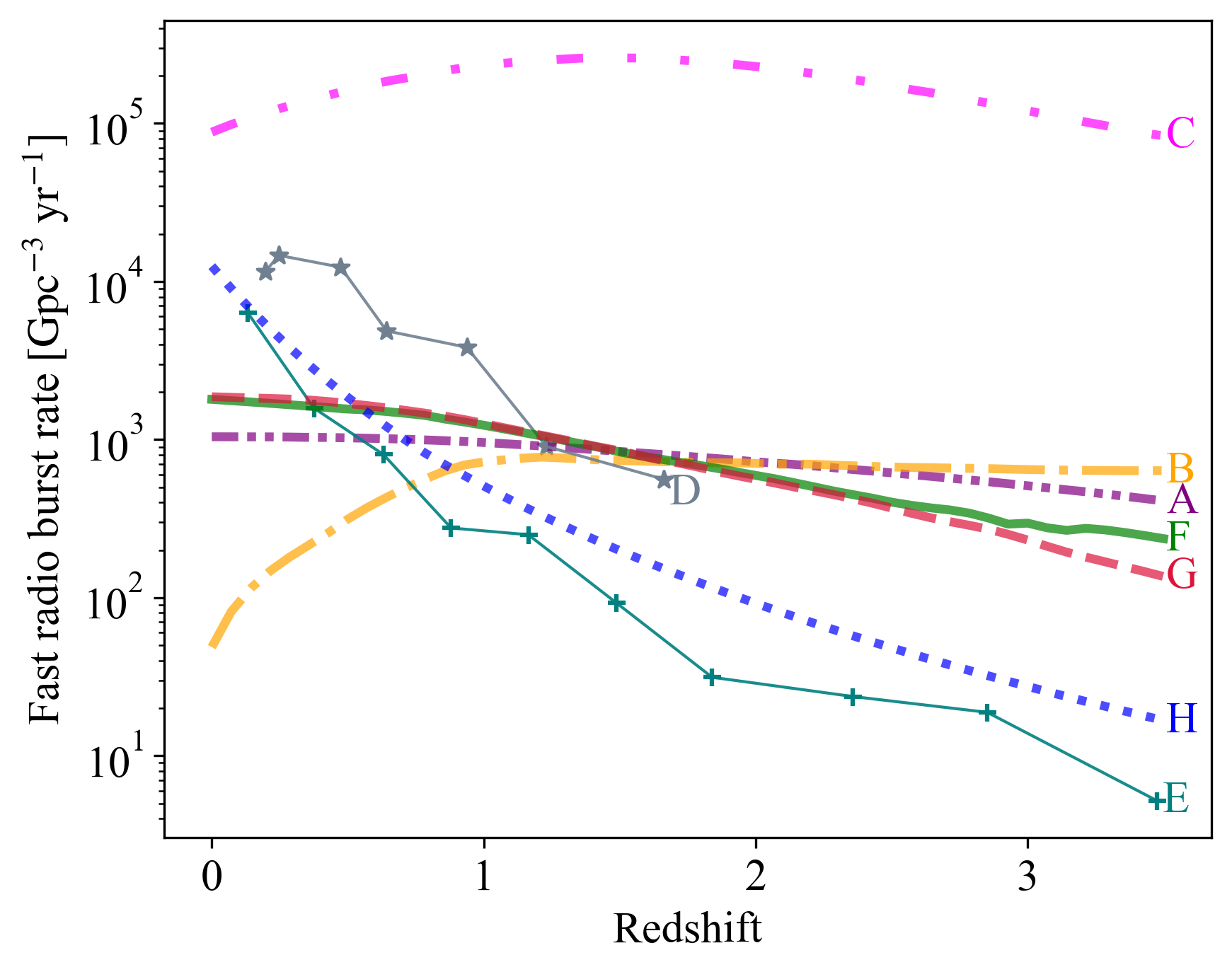}
    \caption{FRB formation/source rates as a function of redshift. \textbf{SFH Models:} \textbf{A:} Accumulated model from \citet{Zhang_Zhang_2022}. \textbf{B:} SFH model from \citet{Zhang_Zhang_2022}. \textbf{C:} SFH model from \citet{Shin_2023}.
    \textbf{Time Delayed Models:} \textbf{D:} Best-fit model of CHIME data from \citet{Hashimoto+2022}. \textbf{E:} Best-fit model of CHIME data from \citet{Chen_2024}. \textbf{F:} Hybrid model that is $20\%$ SFH and $80\%$ lognormal delay with a mean of 13~Gyr from \citet{Zhang_Zhang_2022}.
    \textbf{G:} Lognormal delay model with a central value of 10~Gyr from \citet{Zhang_Zhang_2022}. \textbf{H:} Best-fit model to CHIME data from \citet{Zhang2024}. Here the lognormal delay model, hybrid model, accumulated model and SFH model from \citet{Zhang_Zhang_2022} were scaled up using a multiplicative factor of $1.25 \times 10^4\ {\rm Gpc^{ -3}\,yr^{-1}}$, which equals the value of the FRB rate from \citet{Zhang2024} at redshift $z=0$.}
    \label{fig:frb rates}
\end{figure}

In this section, we briefly summarize results from previous studies on the redshift evolution of FRB rates. It is important here to distinguish what we mean by the FRB rate. There are usually two types considered in the literature: formation/source rate and burst rate. The former is the rate of the formation of the FRB sources. The latter however is the total rate of the occurrence of FRB events. Since a single source could produce repeating FRBs, the former is lower than the latter. It should be noted that all rate models considered in this section describe the source rate of FRBs rather than the burst rate. Although \citet{Zhang_Zhang_2022}, \citet{Hashimoto+2022}, and \citet{Zhang2024} consider both repeaters and non-repeaters in their data, only the first event of the repeater population is used for their analysis. Thus, the rates from these studies can be taken to be source rates. Furthermore, \citet{Chen_2024} excludes the repeater population entirely. Henceforth whenever the FRB rate is mentioned, we are referring to the source rate, unless specified otherwise. The rate models in \citet{Zhang_Zhang_2022} were presented as density functions and hence we re-expressed them to rate units by multiplying by a factor of $1.25\times 10^4\ \text{Gpc}^{-3}\text{yr}^{-1}$ which is the value of the FRB rate in \citet{Zhang2024} at $z=0$. The FRB rate models can be broadly categorized into two types:

\textit{SFH Models:}
The FRB rates as a function of redshift in these models follow the shape of the SFH. We include two such FRB rate curves from \cite{Zhang_Zhang_2022} and \cite{Shin_2023} in \autoref{fig:frb rates}. Also included in this category are models that take the FRB rate to follow the cumulative SFH at any given redshift rather than just the present value (see accumulated model in \citealp{Zhang_Zhang_2022}).

\textit{Time Delayed Models:}
Unlike SFH models, these models take into account the FRB-emitting NSs forming via compact binary coalescences \citep{Hashimoto+2022,Zhang_Zhang_2022,Zhang2024,Chen_2024}. The FRB rates
are obtained by correcting for the coalescence delay time from the SFH model (see lognormal delay model in \citealp{Zhang_Zhang_2022}). This category also includes the hybrid model in \citep{Zhang_Zhang_2022}. We show all relevant models discussed in this section in \autoref{fig:frb rates}.

\subsection{Comparing binary white dwarf coalescence rates with fast radio burst rates}\label{subsec:compare bwd and frb rate}
\begin{figure*}
    \centering
    \includegraphics[width=0.9\textwidth]{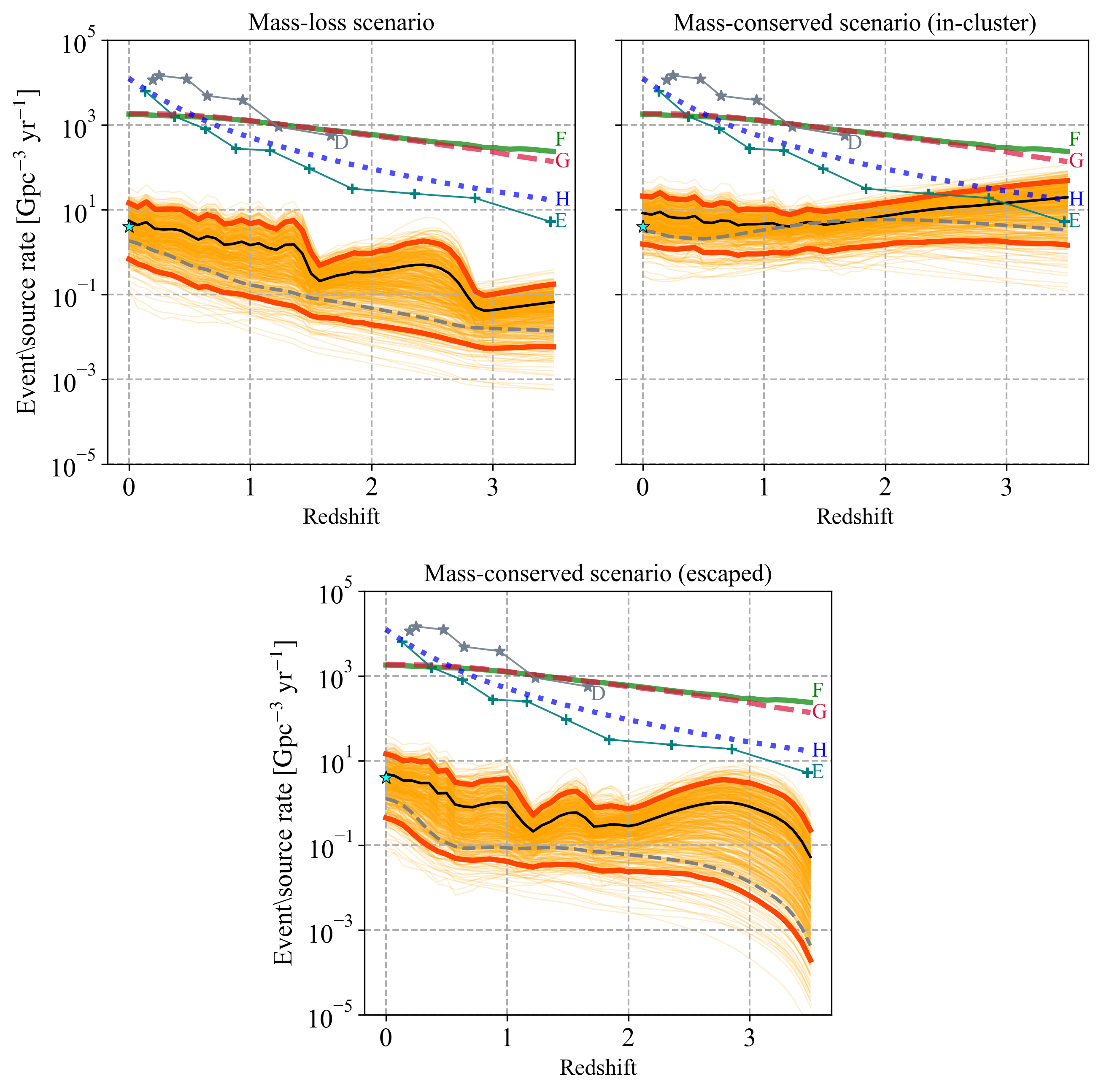}
    \caption{Comparisons between the inferred BWD coalescence rates from GCs outlined in \autoref{subsec:derive bwd rate} and the FRB formation/source rates as a function of redshift. In all the panels, the upper and lower thick solid lines represent the $95^{\text{th}}$ percentile curve and the $5^{\text{th}}$ percentile curve for the BWD coalescence rates, respectively. The solid black curve shows the mean BWD coalescence rate in each panel, and the dashed grey line shows the BWD coalescence rate obtained using the default GC parameters which correspond to a BBH coalescence rate of $10\ \text{Gpc}^{-3}\ \text{yr}^{-1}$ at $z=0$ (see \autoref{subsec:gc formation history}). The legends for the FRB rate curves are as follows. \textbf{D:} Best-fit model of CHIME data from \citet{Hashimoto+2022}. \textbf{E:} Best-fit model of CHIME data from \citet{Chen_2024}. \textbf{F:} Hybrid model that is $20\%$ SFH and $80\%$ lognormal delay with a mean of 13~Gyr from \citet{Zhang_Zhang_2022}.
    \textbf{G:} Lognormal delay model with a central value of 10~Gyr from \citet{Zhang_Zhang_2022}. \textbf{H:} Best-fit model to CHIME data from \citet{Zhang2024}. The cyan star is the rate of BWD coalescences at $z=0$ as determined by \citet{Kremer_2023_2}.
    }
    \label{fig:rate comparison}
\end{figure*}

We compare the obtained BWD coalescence rates from GCs to the various models of the FRB rates as a function of redshift in \autoref{fig:rate comparison}. We note that since this study is concerned with investigating the potential sources of FRBs from BWD coalescences, we will only compare our analysis results with the time-delayed models presented above. Since the analyses on the `mass-loss' and `mass-conserved' scenarios were conducted separately, we present the implications of the rate comparison separately for each case. 

As mentioned in \autoref{subsec:gc simulations}, there are a lot more NS-producing merger events in the `mass-conserved' scenario. This includes many mergers that took place after the BWDs escaped the host GCs. Such escaped mergers are not included in the `mass-loss' scenario since we expect their contribution to be negligible (see also \autoref{fig:bwd delay time dist}). Hence we separate the escaped mergers into a separate set while performing the analysis in the mass-conserved scenario. The results are plotted in \autoref{fig:rate comparison}. All $500$ rate curves obtained from the $500$ samples of $S_{\text{param}}$ are plotted as thin solid lines on the respective panels. We also plot the $5^{\text{th}}$ and the $95^{\text{th}}$ percentile curves (thick solid lines bracketing the thin curves); i.e., the $5^{\text{th}}$ and $95^{\text{th}}$ percentiles of the BWD coalescence rates at each given redshift value. The mean of the rates and the rate obtained by using the default values of $S_{\text{param}}$ are also shown.

Our derived BWD coalescence rates share many similarities to the one derived by \citet{Kremer_2023_2}. The cyan star in each subfigure of \autoref{fig:rate comparison} is the rate of BWD coalescences at $z=0$ as determined by \citet{Kremer_2023_2}. We see that it lies within the uncertainty range of our results in every case. The BWD coalescence rate curve from \citet{Kremer_2023_2} (as shown in Figure~8 of their paper) also shows similar functional forms to our results in the mass-conserved scenario (the adopted scenario in their study), where the coalescence rates at high redshift ($z\approx 3$) is comparable to those at low redshift ($z\approx 0$). The difference between our mass-conserved rate and their rate over redshift is probably due to the different GC formation histories assumed.

In \autoref{tab:rate slopes} we show the power-law exponents of the mean BWD coalescence rate curves (shown in black in \autoref{fig:rate comparison}) between $z=0$ and $z=1$. The exponent were estimated by fitting the mean BWD coalescence rate to a function of $(1+z)^{\alpha}$. This was done for both the mass-loss and mass-conserved scenario. Also shown in \autoref{tab:rate slopes} are the power-law exponents of FRB rate models as presented in their respective studies. In case of the FRB models, two studies (\citealp{Chen_2024} and \citealp{Zhang2024}) provide the analogous power-law exponent. For the studies where the exponents are provided, we directly cite the results. In other cases, we fit the FRB rate curves ourselves and present the slopes.

\subsubsection{The mass-loss scenario}\label{subsubsec:mass loss scenario}
For the mass-loss scenario, the FRB and BWD coalescence rates are shown in the top row of \autoref{fig:rate comparison}. It can be seen that although the FRB rates are about 2-3 orders of magnitude higher than the BWD coalescence rates from GCs, they seem to follow the same general trend with redshift; i.e., there is a steady increase in the rate with decreasing redshift. We present further comparisons to the FRB rate curves for redshifts $z < 1$, since most of the FRB observations to which the models were fit to lie within this redshift range (see discussion presented in Section~7 of \citealp{Shin_2023}). 

It can be seen that the power-law exponents of the FRB rate models in \autoref{tab:rate slopes} are different from the exponents for the mean BWD coalescence rate curves for the mass-loss scenario. In case of \citet{Zhang2024} and \citet{Chen_2024}, the absolute values of exponents are more than twice as large. The exponents of the hybrid and lognormal models from \citet{Zhang_Zhang_2022}, on the other hand, are flatter for $z<1$, with the absolute values of exponents less than half of those of the mean BWD coalescence rate curves. The uncertainty range of the \citet{Hashimoto+2022} model overlaps with those of the mean BWD coalescence rate curves, although the high relative uncertainty implies that we cannot draw any robust conclusions. This high uncertainty is due to the fact that the \citet{Hashimoto+2022} model includes only a few data points compared to the other FRB models.

\begin{table}
    \centering
    \begin{tabular}{|c|c|} \hline 
    \rowcolor{LightCyan}
         \textbf{Model}&  \boldsymbol{$\alpha$}\\ \hline 
    \rowcolor{LightCyan}
 \multicolumn{2}{|c|}{\textbf{Mean BWD coalescence rate curves}}\\\hline
         Mass-loss&  $-1.86\pm 0.16$\\ \hline 
         Mass-conserved (in-cluster)&  $-1.02\pm 0.13$\\ \hline
 Mass-conserved (escaped)& $-2.77\pm 0.25$\\ \hline 
 \rowcolor{LightCyan}
 \multicolumn{2}{|c|}{\textbf{FRB rate models}}\\\hline
         \citet{Zhang2024} model&  $-4.71\pm 0.12$\\ \hline 
 \citet{Chen_2024} model& $-4.90\pm 0.30$\\ \hline
 \citet{Hashimoto+2022} model& $-2.16\pm 0.98$\\ \hline
 \citet{Zhang_Zhang_2022} (hybrid)& $-0.47\pm 0.03$\\ \hline
 \citet{Zhang_Zhang_2022} (lognormal)& $-0.49\pm 0.05$\\ \hline
    \end{tabular}
    \caption{The table lists the values of the power-law exponents of the mean BWD coalescence rates and the FRB rates as a function of $(1+z)^{\alpha}$. For the mean BWD coalescence rates the calculation was restricted to redshifts between $z=0$ and $z=1$. For the FRB rate models the values presented are cited from the respective studies when available and are fit to $(1+z)^{\alpha}$ by ourselves otherwise.}
    \label{tab:rate slopes}
\end{table}

\subsubsection{The mass-conserved scenario}\label{subsubsec:mass conserved scenario}
The comparisons between the BWD coalescence rates derived from the mass-conserved scenario and the FRB models are also shown in \autoref{fig:rate comparison}. The BWD coalescence rates in this case for both in-cluster and escaped events have a different functional dependence to that in the mass-loss limit. The in-cluster rate is roughly flat across redshift. This is in stark contrast to the behaviour of the FRB rate curves which increase with decreasing redshift. The difference in the in-cluster BWD coalescence rate from the BWD coalescence rate in the mass-loss scenario can be attributed to the increase in the number of primordial mergers in the mass-conserved scenario, as was discussed in \autoref{subsec:gc simulations}. The increased number of mergers with short delay times leads to many more merger events at high redshifts as compared to the mass-loss scenario, resulting in flatter rate curves. The coalescence rates for escaped events, however, increase with decreasing redshift, because of the preference for long delay times. This is quite similar to the trend in the mass-loss scenario, because the delay time distributions are similar in the two cases (see \autoref{fig:bwd delay time dist}).

For the mass-conserved scenario, we present the power-law slopes (for redshifts between $z=0$ and $z=1$) separately for in-cluster and out-of-cluster events (see \autoref{tab:rate slopes}). The exponent for the in-cluster BWD coalescence rate does not seem to agree with the exponents of the FRB rate models from the \citet{Zhang2024}, \citet{Chen_2024} and \citet{Zhang_Zhang_2022} models. However, the power-law slopes for the escaped BWD coalescence rates are much steeper than that for the in-cluster rates (for $z<1$). This brings them much closer in similarity to the \citet{Hashimoto+2022} FRB models as seen in \autoref{tab:rate slopes}. As mentioned before, we cannot draw robust conclusions regarding the \citet{Hashimoto+2022} FRB model due to the high relative uncertainty in its power law exponent.

\section{Discussion}\label{sec:discussion}
\subsection{Uncertainties in the globular cluster formation history}\label{subsec:gc formation history uncertainty}
One of the main uncertainties in the BWD coalescence rates comes from the uncertain initial GC property distributions and formation histories (see \autoref{subsec:gc formation history}). In this study, we use GC population constraints derived from the rate of BBH coalescences that GCs could produce. Although the local rate of BBH coalescences is well-measured from GW observations, the inferred rate of BBH coalescences originating from GCs across cosmic time is as a function of spin and redshift, which carries large uncertainty. This uncertainty propagates to the inferred parameters governing the GC population, leading to broad posteriors on the parameters (see Figure~B1 in \citealp{Fishbach_Fragione_2023}). Thus, our predicted BWD coalescence rates over redshift also show a large spread (\autoref{fig:rate comparison}). Future studies, using GWs as well as other probes of GCs, will better constrain the posterior distributions of these parameters, providing tighter predictions for the BWD coalescence rate from GCs as a function of redshift. Although the spread of the BWD coalescence rate is poorly constrained, it can be seen from \autoref{fig:rate comparison} that the functional form of the curves is similar in each case. Hence, we are still able to get a robust comparison of the functional forms of the BWD coalescence rates and the FRB rates.

\newpage
\subsection{Uncertainties in binary white dwarf coalescences}\label{subsec: bwd coalescence uncertainty}
The uncertainties in the mass accretion efficiency and mass-loss rate for BWD mergers are bracketed by the two different scenarios in this study, mass-loss and mass-conserved. These scenarios have significant impacts on both the number of FRB-emitting NSs formed in GCs and their redshift evolution, as shown in \autoref{fig:rate comparison}. For instance, in the mass-conserved scenario, the BWD coalescence rates from GCs at higher redshift ($\gtrsim 3$) can be comparable to or even larger than the observed FRB rates due to the increased number of coalescences from BWDs formed in primordial binaries. Therefore, FRB observations at higher redshift, combining with the inferred GC formation history from GW detections, may be used to test the mechanism of BWD coalescences.

Similarly, BWD coalescences in galactic fields may also produce FRB-emitting NSs. The coalescence rate of two super-Chandrasekhar CO WDs in the field is roughly an order of magnitude lower than the rates of Type Ia supernovae ($\sim 10^4\,{\rm Gpc^{-3}\,yr^{-1}}$; \citealp[e.g.,][]{Badenes_Maoz_2012,Maoz_Mannucci_2012}), which is close to the FRB source rates in the local Universe. However, \citet{Ruiter+2009} showed that the delay time of BWD coalescences in the field is about 1~Gyr, similar to the delay time of the cluster BWD mergers from primordial binaries ($\sim 100~$Myr). Therefore, the redshift evolution of field BWD coalescences likely follows the SFH more closely than old stellar populations. This suggests that if BWD coalescences from binary evolution are indeed one of the dominant mechanism for producing FRB sources both in the field and in clusters, the FRB formation rates would likely peak at higher redshift rather than decreasing with increasing redshift.

While we have fixed the initial binary fraction to 5\% in our cluster simulations, we do not expect the results to be affected significantly by this choice. The number of BBH mergers is mostly dominated by the dynamically-produced systems \citep[e.g.,][]{Chatterjee+2017}. A larger initial binary fraction likely also has negligible effects on the number of BWD mergers \citep[see, e.g.,][their Section~7.3]{Kremer2021}. Increasing the initial binary fraction from $5\%$ to $50\%$ and varying the initial semi-major axis distribution, for example, will likely increase the number of BWD mergers by a factor of 10 at most, which does not affect the conclusions of this study.

\section{Conclusions}\label{sec:conclusion}
In this study, we predicted the rate of BWD coalescences resulting in potentially FRB-emitting NSs in GCs as a function of redshift. This was done by using the data on BWD coalescences from \texttt{CMC} simulations of realistic GCs \citep{Kremer_2020}. We used GW observations of BBH coalescences \citep{Fishbach_Fragione_2023} to constrain the GC formation histories and property distributions. We compared the resulting BWD coalescence rates to various FRB source rate models from the literature, most of which find that the measured FRB source rate does not track the SFH but instead the old stellar populations. Hence, we explored alternative explanations such as NSs from the collisions and mergers of WDs which were proposed as a potential source of FRBs.

In the mass-loss scenario, the BWD coalescence rates seem to track the FRB rate models in terms of functional dependence with redshift. Both rates increase with decreasing redshift. In the mass-conserved scenario, the BWD coalescences occurring within GCs are almost redshift independent, and BWD coalescence rates are comparable to the FRB source formation rates at high redshift ($z\approx 3$). The BWD coalescence rates from escaped binaries, on the other hand, follow the same trend with redshift as the FRB rates. 

Overall, if all NSs formed from BWD coalescences produce FRBs, GCs would contribute to $\lesssim 1\%$ of the total FRB sources in the local Universe ($z<1$) for any BWD merger mass scenario, consistent with previous studies \citep{Bhardwaj+2024,Bhardwaj+2024_hostgalaxy}. According to the \texttt{CMC} simulations, GCs cannot provide a significantly larger contribution to the FRB rate without overpredicting the BBH merger rate. NSs produced through BWD coalescences in the galactic fields may account for the majority of the FRB source formation rates in the local Universe \citep[e.g.,][]{Badenes_Maoz_2012,Maoz_Mannucci_2012}. However, due to the short delay times of field BWD coalescences \citep[e.g.,][]{Ruiter+2009}, FRB source rates would closely trace the SFH if BWD coalescence is the dominant formation channel. Similarly if magnetars from core-collapse supernovae is the dominant formation channel, the above is true.

The CHIME array will be expanded with additional outrigger telescopes, enhancing its ability to localize FRB detections more effectively. The first of these outriggers, named the k’niPatn k’l stk’masqt outrigger (KKO) telescope, has been built at the time of writing with two more to be commissioned \citep{KKO_chime}. With better FRB localizations resulting from these outriggers, we will be able to constrain the formation site of FRB sources and refine the current FRB rate models. Meanwhile, planned upgrades to the LVK observatory network will lead to thousands of new BBH detections out to redshifts $z\approx2$ in the next few years, with implications for the GC population. Galaxy observations are independently improving our knowledge of GC formation histories \citep{Kissler_2000,Forbes_2018}. Our work in this study allow us to better interpret these future FRB, GW, and GC observations. 

\begin{acknowledgments}
    We are grateful to Kyle Kremer and Mohit Bhardwaj for useful discussions and to the anonymous referee for helpful comments. C.S.Y. acknowledges support from the Natural Sciences and Engineering Research Council of Canada (NSERC) DIS-2022-568580. M.F. acknowledges support from NSERC RGPIN-2023-05511, the University of Toronto Connaught Fund, and the Alfred P. Sloan Foundation.
\end{acknowledgments}

\bibliography{BWD_FRBs}

\begin{thebibliography}{}
\expandafter\ifx\csname natexlab\endcsname\relax\def\natexlab#1{#1}\fi
\providecommand{\url}[1]{\href{#1}{#1}}
\providecommand{\dodoi}[1]{doi:~\href{http://doi.org/#1}{\nolinkurl{#1}}}
\providecommand{\doeprint}[1]{\href{http://ascl.net/#1}{\nolinkurl{http://ascl.net/#1}}}
\providecommand{\doarXiv}[1]{\href{https://arxiv.org/abs/#1}{\nolinkurl{https://arxiv.org/abs/#1}}}

\bibitem[{{Badenes} \& {Maoz}(2012)}]{Badenes_Maoz_2012}
{Badenes}, C., \& {Maoz}, D. 2012, \apjl, 749, L11, \dodoi{10.1088/2041-8205/749/1/L11}

\bibitem[{{Bhardwaj} {et~al.}(2024{\natexlab{a}}){Bhardwaj}, {Lee}, \& {Ji}}]{Bhardwaj+2024}
{Bhardwaj}, M., {Lee}, J., \& {Ji}, K. 2024{\natexlab{a}}, \nat, 634, 1065, \dodoi{10.1038/s41586-024-08065-w}

\bibitem[{{Bhardwaj} {et~al.}(2021){Bhardwaj}, {Gaensler}, {Kaspi}, {Landecker}, {Mckinven}, {Michilli}, {Pleunis}, {Tendulkar}, {Andersen}, {Boyle}, {Cassanelli}, {Chawla}, {Cook}, {Dobbs}, {Fonseca}, {Kaczmarek}, {Leung}, {Masui}, {Mnchmeyer}, {Ng}, {Rafiei-Ravandi}, {Scholz}, {Shin}, {Smith}, {Stairs}, \& {Zwaniga}}]{Bhardwaj_2021}
{Bhardwaj}, M., {Gaensler}, B.~M., {Kaspi}, V.~M., {et~al.} 2021, \apjl, 910, L18, \dodoi{10.3847/2041-8213/abeaa6}

\bibitem[{{Bhardwaj} {et~al.}(2024{\natexlab{b}}){Bhardwaj}, {Michilli}, {Kirichenko}, {Modilim}, {Shin}, {Kaspi}, {Andersen}, {Cassanelli}, {Brar}, {Chatterjee}, {Cook}, {Dong}, {Fonseca}, {Gaensler}, {Ibik}, {Kaczmarek}, {Lanman}, {Leung}, {Masui}, {Pandhi}, {Pearlman}, {Petroff}, {Pleunis}, {Prochaska}, {Rafiei-Ravandi}, {Sand}, {Scholz}, \& {Smith}}]{Bhardwaj+2024_hostgalaxy}
{Bhardwaj}, M., {Michilli}, D., {Kirichenko}, A.~Y., {et~al.} 2024{\natexlab{b}}, \apjl, 971, L51, \dodoi{10.3847/2041-8213/ad64d1}

\bibitem[{{Bochenek} {et~al.}(2020){Bochenek}, {Ravi}, {Belov}, {Hallinan}, {Kocz}, {Kulkarni}, \& {McKenna}}]{galactic_magnetar}
{Bochenek}, C.~D., {Ravi}, V., {Belov}, K.~V., {et~al.} 2020, \nat, 587, 59, \dodoi{10.1038/s41586-020-2872-x}

\bibitem[{{Breivik} {et~al.}(2020){Breivik}, {Coughlin}, {Zevin}, {Rodriguez}, {Kremer}, {Ye}, {Andrews}, {Kurkowski}, {Digman}, {Larson}, \& {Rasio}}]{cosmic}
{Breivik}, K., {Coughlin}, S., {Zevin}, M., {et~al.} 2020, \apj, 898, 71, \dodoi{10.3847/1538-4357/ab9d85}

\bibitem[{{Burke-Spolaor} \& {Bannister}(2014)}]{Burke_2014}
{Burke-Spolaor}, S., \& {Bannister}, K.~W. 2014, \apj, 792, 19, \dodoi{10.1088/0004-637X/792/1/19}

\bibitem[{{Champion} {et~al.}(2016){Champion}, {Petroff}, {Kramer}, {Keith}, {Bailes}, {Barr}, {Bates}, {Bhat}, {Burgay}, {Burke-Spolaor}, {Flynn}, {Jameson}, {Johnston}, {Ng}, {Levin}, {Possenti}, {Stappers}, {van Straten}, {Thornton}, {Tiburzi}, \& {Lyne}}]{Champion_2016}
{Champion}, D.~J., {Petroff}, E., {Kramer}, M., {et~al.} 2016, \mnras, 460, L30, \dodoi{10.1093/mnrasl/slw069}

\bibitem[{{Chatterjee} {et~al.}(2017){Chatterjee}, {Rodriguez}, \& {Rasio}}]{Chatterjee+2017}
{Chatterjee}, S., {Rodriguez}, C.~L., \& {Rasio}, F.~A. 2017, \apj, 834, 68, \dodoi{10.3847/1538-4357/834/1/68}

\bibitem[{{Chen} {et~al.}(2024){Chen}, {Jia}, {Dong}, \& {Wang}}]{Chen_2024}
{Chen}, J.~H., {Jia}, X.~D., {Dong}, X.~F., \& {Wang}, F.~Y. 2024, \apjl, 973, L54, \dodoi{10.3847/2041-8213/ad7b39}

\bibitem[{{CHIME/FRB Collaboration} {et~al.}(2020){CHIME/FRB Collaboration}, {Andersen}, {Bandura}, {Bhardwaj}, {Bij}, {Boyce}, {Boyle}, {Brar}, {Cassanelli}, {Chawla}, {Chen}, {Cliche}, {Cook}, {Cubranic}, {Curtin}, {Denman}, {Dobbs}, {Dong}, {Fandino}, {Fonseca}, {Gaensler}, {Giri}, {Good}, {Halpern}, {Hill}, {Hinshaw}, {H{\"o}fer}, {Josephy}, {Kania}, {Kaspi}, {Landecker}, {Leung}, {Li}, {Lin}, {Masui}, {McKinven}, {Mena-Parra}, {Merryfield}, {Meyers}, {Michilli}, {Milutinovic}, {Mirhosseini}, {M{\"u}nchmeyer}, {Naidu}, {Newburgh}, {Ng}, {Patel}, {Pen}, {Pinsonneault-Marotte}, {Pleunis}, {Quine}, {Rafiei-Ravandi}, {Rahman}, {Ransom}, {Renard}, {Sanghavi}, {Scholz}, {Shaw}, {Shin}, {Siegel}, {Singh}, {Smegal}, {Smith}, {Stairs}, {Tan}, {Tendulkar}, {Tretyakov}, {Vanderlinde}, {Wang}, {Wulf}, \& {Zwaniga}}]{CHIME_GalacticFRB}
{CHIME/FRB Collaboration}, {Andersen}, B.~C., {Bandura}, K.~M., {et~al.} 2020, \nat, 587, 54, \dodoi{10.1038/s41586-020-2863-y}

\bibitem[{{CHIME/FRB Collaboration} {et~al.}(2021){CHIME/FRB Collaboration}, {Amiri}, {Andersen}, {Bandura}, {Berger}, {Bhardwaj}, {Boyce}, {Boyle}, {Brar}, {Breitman}, {Cassanelli}, {Chawla}, {Chen}, {Cliche}, {Cook}, {Cubranic}, {Curtin}, {Deng}, {Dobbs}, {Dong}, {Eadie}, {Fandino}, {Fonseca}, {Gaensler}, {Giri}, {Good}, {Halpern}, {Hill}, {Hinshaw}, {Josephy}, {Kaczmarek}, {Kader}, {Kania}, {Kaspi}, {Landecker}, {Lang}, {Leung}, {Li}, {Lin}, {Masui}, {McKinven}, {Mena-Parra}, {Merryfield}, {Meyers}, {Michilli}, {Milutinovic}, {Mirhosseini}, {M{\"u}nchmeyer}, {Naidu}, {Newburgh}, {Ng}, {Patel}, {Pen}, {Petroff}, {Pinsonneault-Marotte}, {Pleunis}, {Rafiei-Ravandi}, {Rahman}, {Ransom}, {Renard}, {Sanghavi}, {Scholz}, {Shaw}, {Shin}, {Siegel}, {Sikora}, {Singh}, {Smith}, {Stairs}, {Tan}, {Tendulkar}, {Vanderlinde}, {Wang}, {Wulf}, \& {Zwaniga}}]{CHIME_Amiri_2021}
{CHIME/FRB Collaboration}, {Amiri}, M., {Andersen}, B.~C., {et~al.} 2021, \apjs, 257, 59, \dodoi{10.3847/1538-4365/ac33ab}

\bibitem[{{Duquennoy} \& {Mayor}(1991)}]{Duquennoy_Mayor_1991}
{Duquennoy}, A., \& {Mayor}, M. 1991, \aap, 248, 485

\bibitem[{{Fishbach} \& {Fragione}(2023)}]{Fishbach_Fragione_2023}
{Fishbach}, M., \& {Fragione}, G. 2023, \mnras, 522, 5546, \dodoi{10.1093/mnras/stad1364}

\bibitem[{{Forbes} {et~al.}(2018){Forbes}, {Read}, {Gieles}, \& {Collins}}]{Forbes_2018}
{Forbes}, D.~A., {Read}, J.~I., {Gieles}, M., \& {Collins}, M. L.~M. 2018, \mnras, 481, 5592, \dodoi{10.1093/mnras/sty2584}

\bibitem[{{Gratton} {et~al.}(2019){Gratton}, {Bragaglia}, {Carretta}, {D'Orazi}, {Lucatello}, \& {Sollima}}]{Gratton_2019}
{Gratton}, R., {Bragaglia}, A., {Carretta}, E., {et~al.} 2019, \aapr, 27, 8, \dodoi{10.1007/s00159-019-0119-3}

\bibitem[{{Hashimoto} {et~al.}(2022){Hashimoto}, {Goto}, {Chen}, {Ho}, {Hsiao}, {Wong}, {On}, {Kim}, {Kilerci-Eser}, {Huang}, {Santos}, \& {Yamasaki}}]{Hashimoto+2022}
{Hashimoto}, T., {Goto}, T., {Chen}, B.~H., {et~al.} 2022, \mnras, 511, 1961, \dodoi{10.1093/mnras/stac065}

\bibitem[{{James} {et~al.}(2022){James}, {Prochaska}, {Macquart}, {North-Hickey}, {Bannister}, \& {Dunning}}]{James2022}
{James}, C.~W., {Prochaska}, J.~X., {Macquart}, J.~P., {et~al.} 2022, \mnras, 510, L18, \dodoi{10.1093/mnrasl/slab117}

\bibitem[{{Kashiyama} {et~al.}(2013){Kashiyama}, {Ioka}, \& {M{\'e}sz{\'a}ros}}]{Kashiyama+2013}
{Kashiyama}, K., {Ioka}, K., \& {M{\'e}sz{\'a}ros}, P. 2013, \apjl, 776, L39, \dodoi{10.1088/2041-8205/776/2/L39}

\bibitem[{{Kaspi} \& {Beloborodov}(2017)}]{Kaspi_2017}
{Kaspi}, V.~M., \& {Beloborodov}, A.~M. 2017, \araa, 55, 261, \dodoi{10.1146/annurev-astro-081915-023329}

\bibitem[{{Keane} {et~al.}(2012){Keane}, {Stappers}, {Kramer}, \& {Lyne}}]{Keane_2012}
{Keane}, E.~F., {Stappers}, B.~W., {Kramer}, M., \& {Lyne}, A.~G. 2012, \mnras, 425, L71, \dodoi{10.1111/j.1745-3933.2012.01306.x}

\bibitem[{{Kirsten} {et~al.}(2022){Kirsten}, {Marcote}, {Nimmo}, {Hessels}, {Bhardwaj}, {Tendulkar}, {Keimpema}, {Yang}, {Snelders}, {Scholz}, {Pearlman}, {Law}, {Peters}, {Giroletti}, {Paragi}, {Bassa}, {Hewitt}, {Bach}, {Bezrukovs}, {Burgay}, {Buttaccio}, {Conway}, {Corongiu}, {Feiler}, {Forss{\'e}n}, {Gawro{\'n}ski}, {Karuppusamy}, {Kharinov}, {Lindqvist}, {Maccaferri}, {Melnikov}, {Ould-Boukattine}, {Possenti}, {Surcis}, {Wang}, {Yuan}, {Aggarwal}, {Anna-Thomas}, {Bower}, {Blaauw}, {Burke-Spolaor}, {Cassanelli}, {Clarke}, {Fonseca}, {Gaensler}, {Gopinath}, {Kaspi}, {Kassim}, {Lazio}, {Leung}, {Li}, {Lin}, {Masui}, {Mckinven}, {Michilli}, {Mikhailov}, {Ng}, {Orbidans}, {Pen}, {Petroff}, {Rahman}, {Ransom}, {Shin}, {Smith}, {Stairs}, \& {Vlemmings}}]{Kirsten_2022}
{Kirsten}, F., {Marcote}, B., {Nimmo}, K., {et~al.} 2022, \nat, 602, 585, \dodoi{10.1038/s41586-021-04354-w}

\bibitem[{{Kissler-Patig}(2000)}]{Kissler_2000}
{Kissler-Patig}, M. 2000, Reviews in Modern Astronomy, 13, 13, \dodoi{10.48550/arXiv.astro-ph/0002070}

\bibitem[{{Kremer} {et~al.}(2023{\natexlab{a}}){Kremer}, {Fuller}, {Piro}, \& {Ransom}}]{Kremer+2023}
{Kremer}, K., {Fuller}, J., {Piro}, A.~L., \& {Ransom}, S.~M. 2023{\natexlab{a}}, \mnras, 525, L22, \dodoi{10.1093/mnrasl/slad088}

\bibitem[{{Kremer} {et~al.}(2023{\natexlab{b}}){Kremer}, {Li}, {Lu}, {Piro}, \& {Zhang}}]{Kremer_2023_2}
{Kremer}, K., {Li}, D., {Lu}, W., {Piro}, A.~L., \& {Zhang}, B. 2023{\natexlab{b}}, \apj, 944, 6, \dodoi{10.3847/1538-4357/acabbf}

\bibitem[{{Kremer} {et~al.}(2021{\natexlab{a}}){Kremer}, {Piro}, \& {Li}}]{Kremer+2021_frb}
{Kremer}, K., {Piro}, A.~L., \& {Li}, D. 2021{\natexlab{a}}, \apjl, 917, L11, \dodoi{10.3847/2041-8213/ac13a0}

\bibitem[{{Kremer} {et~al.}(2021{\natexlab{b}}){Kremer}, {Rui}, {Weatherford}, {Chatterjee}, {Fragione}, {Rasio}, {Rodriguez}, \& {Ye}}]{Kremer2021}
{Kremer}, K., {Rui}, N.~Z., {Weatherford}, N.~C., {et~al.} 2021{\natexlab{b}}, \apj, 917, 28, \dodoi{10.3847/1538-4357/ac06d4}

\bibitem[{{Kremer} {et~al.}(2020{\natexlab{a}}){Kremer}, {Ye}, {Chatterjee}, {Rodriguez}, \& {Rasio}}]{bh_burning}
{Kremer}, K., {Ye}, C.~S., {Chatterjee}, S., {Rodriguez}, C.~L., \& {Rasio}, F.~A. 2020{\natexlab{a}}, in IAU Symposium, Vol. 351, Star Clusters: From the Milky Way to the Early Universe, ed. A.~{Bragaglia}, M.~{Davies}, A.~{Sills}, \& E.~{Vesperini}, 357--366, \dodoi{10.1017/S1743921319007269}

\bibitem[{{Kremer} {et~al.}(2020{\natexlab{b}}){Kremer}, {Ye}, {Rui}, {Weatherford}, {Chatterjee}, {Fragione}, {Rodriguez}, {Spera}, \& {Rasio}}]{Kremer_2020}
{Kremer}, K., {Ye}, C.~S., {Rui}, N.~Z., {et~al.} 2020{\natexlab{b}}, \apjs, 247, 48, \dodoi{10.3847/1538-4365/ab7919}

\bibitem[{{Kroupa}(2001)}]{Kroupa_2001}
{Kroupa}, P. 2001, \mnras, 322, 231, \dodoi{10.1046/j.1365-8711.2001.04022.x}

\bibitem[{{Lanman} {et~al.}(2024){Lanman}, {Andrew}, {Lazda}, {Shah}, {Amiri}, {Balasubramanian}, {Bandura}, {Boyle}, {Brar}, {Carlson}, {Cliche}, {Gusinskaia}, {Hendricksen}, {Kaczmarek}, {Landecker}, {Leung}, {Mckinven}, {Mena-Parra}, {Milutinovic}, {Nimmo}, {Pearlman}, {Renard}, {Rahman}, {Shaw}, {Siegel}, {Smegal}, {Cassanelli}, {Chatterjee}, {Curtin}, {Dobbs}, {Dong}, {Halpern}, {Hopkins}, {Kaspi}, {Khairy}, {Masui}, {Meyers}, {Michilli}, {Petroff}, {Pinsonneault-Marotte}, {Pleunis}, {Rafiei-Ravandi}, {Shin}, {Smith}, {Vanderlinde}, \& {Zegmott}}]{KKO_chime}
{Lanman}, A.~E., {Andrew}, S., {Lazda}, M., {et~al.} 2024, \aj, 168, 87, \dodoi{10.3847/1538-3881/ad5838}

\bibitem[{{Law} {et~al.}(2024){Law}, {Sharma}, {Ravi}, {Chen}, {Catha}, {Connor}, {Faber}, {Hallinan}, {Harnach}, {Hellbourg}, {Hobbs}, {Hodge}, {Hodges}, {Lamb}, {Rasmussen}, {Sherman}, {Shi}, {Simard}, {Squillace}, {Weinreb}, {Woody}, \& {Yurk}}]{Law+2024}
{Law}, C.~J., {Sharma}, K., {Ravi}, V., {et~al.} 2024, \apj, 967, 29, \dodoi{10.3847/1538-4357/ad3736}

\bibitem[{Lorimer {et~al.}(2007)Lorimer, Bailes, McLaughlin, Narkevic, \& Crawford}]{Lorimer_2007}
Lorimer, D.~R., Bailes, M., McLaughlin, M.~A., Narkevic, D.~J., \& Crawford, F. 2007, Science, 318, 777, \dodoi{10.1126/science.1147532}

\bibitem[{{Lu} {et~al.}(2022){Lu}, {Beniamini}, \& {Kumar}}]{Lu+2022}
{Lu}, W., {Beniamini}, P., \& {Kumar}, P. 2022, \mnras, 510, 1867, \dodoi{10.1093/mnras/stab3500}

\bibitem[{{Madau} \& {Dickinson}(2014)}]{Madau_Dickinson_2014}
{Madau}, P., \& {Dickinson}, M. 2014, \araa, 52, 415, \dodoi{10.1146/annurev-astro-081811-125615}

\bibitem[{{Madau} \& {Fragos}(2017)}]{Mandau_Fragos_2017}
{Madau}, P., \& {Fragos}, T. 2017, \apj, 840, 39, \dodoi{10.3847/1538-4357/aa6af9}

\bibitem[{{Maoz} \& {Mannucci}(2012)}]{Maoz_Mannucci_2012}
{Maoz}, D., \& {Mannucci}, F. 2012, \pasa, 29, 447, \dodoi{10.1071/AS11052}

\bibitem[{{Marcote} {et~al.}(2017){Marcote}, {Paragi}, {Hessels}, {Keimpema}, {van Langevelde}, {Huang}, {Bassa}, {Bogdanov}, {Bower}, {Burke-Spolaor}, {Butler}, {Campbell}, {Chatterjee}, {Cordes}, {Demorest}, {Garrett}, {Ghosh}, {Kaspi}, {Law}, {Lazio}, {McLaughlin}, {Ransom}, {Salter}, {Scholz}, {Seymour}, {Siemion}, {Spitler}, {Tendulkar}, \& {Wharton}}]{Marcote_2017}
{Marcote}, B., {Paragi}, Z., {Hessels}, J.~W.~T., {et~al.} 2017, \apjl, 834, L8, \dodoi{10.3847/2041-8213/834/2/L8}

\bibitem[{{Masui} {et~al.}(2015){Masui}, {Lin}, {Sievers}, {Anderson}, {Chang}, {Chen}, {Ganguly}, {Jarvis}, {Kuo}, {Li}, {Liao}, {McLaughlin}, {Pen}, {Peterson}, {Roman}, {Timbie}, {Voytek}, \& {Yadav}}]{gbt_2015}
{Masui}, K., {Lin}, H.-H., {Sievers}, J., {et~al.} 2015, \nat, 528, 523, \dodoi{10.1038/nature15769}

\bibitem[{{Petroff} {et~al.}(2022){Petroff}, {Hessels}, \& {Lorimer}}]{Petroff_2022}
{Petroff}, E., {Hessels}, J.~W.~T., \& {Lorimer}, D.~R. 2022, \aapr, 30, 2, \dodoi{10.1007/s00159-022-00139-w}

\bibitem[{{Petroff} {et~al.}(2015){Petroff}, {Bailes}, {Barr}, {Barsdell}, {Bhat}, {Bian}, {Burke-Spolaor}, {Caleb}, {Champion}, {Chandra}, {Da Costa}, {Delvaux}, {Flynn}, {Gehrels}, {Greiner}, {Jameson}, {Johnston}, {Kasliwal}, {Keane}, {Keller}, {Kocz}, {Kramer}, {Leloudas}, {Malesani}, {Mulchaey}, {Ng}, {Ofek}, {Perley}, {Possenti}, {Schmidt}, {Shen}, {Stappers}, {Tisserand}, {van Straten}, \& {Wolf}}]{Petroff_2015}
{Petroff}, E., {Bailes}, M., {Barr}, E.~D., {et~al.} 2015, \mnras, 447, 246, \dodoi{10.1093/mnras/stu2419}

\bibitem[{{Planck Collaboration} {et~al.}(2020){Planck Collaboration}, {Aghanim, N.}, {Akrami, Y.}, {Ashdown, M.}, {Aumont, J.}, {Baccigalupi, C.}, {Ballardini, M.}, {Banday, A. J.}, {Barreiro, R. B.}, {Bartolo, N.}, {Basak, S.}, {Battye, R.}, {Benabed, K.}, {Bernard, J.-P.}, {Bersanelli, M.}, {Bielewicz, P.}, {Bock, J. J.}, {Bond, J. R.}, {Borrill, J.}, {Bouchet, F. R.}, {Boulanger, F.}, {Bucher, M.}, {Burigana, C.}, {Butler, R. C.}, {Calabrese, E.}, {Cardoso, J.-F.}, {Carron, J.}, {Challinor, A.}, {Chiang, H. C.}, {Chluba, J.}, {Colombo, L. P. L.}, {Combet, C.}, {Contreras, D.}, {Crill, B. P.}, {Cuttaia, F.}, {de Bernardis, P.}, {de Zotti, G.}, {Delabrouille, J.}, {Delouis, J.-M.}, {Di Valentino, E.}, {Diego, J. M.}, {Dor\'e, O.}, {Douspis, M.}, {Ducout, A.}, {Dupac, X.}, {Dusini, S.}, {Efstathiou, G.}, {Elsner, F.}, {En\ss{}lin, T. A.}, {Eriksen, H. K.}, {Fantaye, Y.}, {Farhang, M.}, {Fergusson, J.}, {Fernandez-Cobos, R.}, {Finelli, F.}, {Forastieri, F.}, {Frailis, M.}, {Fraisse, A. A.}, {Franceschi, E.},
  {Frolov, A.}, {Galeotta, S.}, {Galli, S.}, {Ganga, K.}, {G\'enova-Santos, R. T.}, {Gerbino, M.}, {Ghosh, T.}, {Gonz\'alez-Nuevo, J.}, {G\'orski, K. M.}, {Gratton, S.}, {Gruppuso, A.}, {Gudmundsson, J. E.}, {Hamann, J.}, {Handley, W.}, {Hansen, F. K.}, {Herranz, D.}, {Hildebrandt, S. R.}, {Hivon, E.}, {Huang, Z.}, {Jaffe, A. H.}, {Jones, W. C.}, {Karakci, A.}, {Keih\"anen, E.}, {Keskitalo, R.}, {Kiiveri, K.}, {Kim, J.}, {Kisner, T. S.}, {Knox, L.}, {Krachmalnicoff, N.}, {Kunz, M.}, {Kurki-Suonio, H.}, {Lagache, G.}, {Lamarre, J.-M.}, {Lasenby, A.}, {Lattanzi, M.}, {Lawrence, C. R.}, {Le Jeune, M.}, {Lemos, P.}, {Lesgourgues, J.}, {Levrier, F.}, {Lewis, A.}, {Liguori, M.}, {Lilje, P. B.}, {Lilley, M.}, {Lindholm, V.}, {L\'opez-Caniego, M.}, {Lubin, P. M.}, {Ma, Y.-Z.}, {Mac\'{\i}as-P\'erez, J. F.}, {Maggio, G.}, {Maino, D.}, {Mandolesi, N.}, {Mangilli, A.}, {Marcos-Caballero, A.}, {Maris, M.}, {Martin, P. G.}, {Martinelli, M.}, {Mart\'{\i}nez-Gonz\'alez, E.}, {Matarrese, S.}, {Mauri, N.}, {McEwen, J. D.},
  {Meinhold, P. R.}, {Melchiorri, A.}, {Mennella, A.}, {Migliaccio, M.}, {Millea, M.}, {Mitra, S.}, {Miville-Desch\^enes, M.-A.}, {Molinari, D.}, {Montier, L.}, {Morgante, G.}, {Moss, A.}, {Natoli, P.}, {N\o{}rgaard-Nielsen, H. U.}, {Pagano, L.}, {Paoletti, D.}, {Partridge, B.}, {Patanchon, G.}, {Peiris, H. V.}, {Perrotta, F.}, {Pettorino, V.}, {Piacentini, F.}, {Polastri, L.}, {Polenta, G.}, {Puget, J.-L.}, {Rachen, J. P.}, {Reinecke, M.}, {Remazeilles, M.}, {Renzi, A.}, {Rocha, G.}, {Rosset, C.}, {Roudier, G.}, {Rubi\~no-Mart\'{\i}n, J. A.}, {Ruiz-Granados, B.}, {Salvati, L.}, {Sandri, M.}, {Savelainen, M.}, {Scott, D.}, {Shellard, E. P. S.}, {Sirignano, C.}, {Sirri, G.}, {Spencer, L. D.}, {Sunyaev, R.}, {Suur-Uski, A.-S.}, {Tauber, J. A.}, {Tavagnacco, D.}, {Tenti, M.}, {Toffolatti, L.}, {Tomasi, M.}, {Trombetti, T.}, {Valenziano, L.}, {Valiviita, J.}, {Van Tent, B.}, {Vibert, L.}, {Vielva, P.}, {Villa, F.}, {Vittorio, N.}, {Wandelt, B. D.}, {Wehus, I. K.}, {White, M.}, {White, S. D. M.}, {Zacchei, A.}, \&
  {Zonca, A.}}]{Planck_2018}
{Planck Collaboration}, {Aghanim, N.}, {Akrami, Y.}, {et~al.} 2020, A\&A, 641, A6, \dodoi{10.1051/0004-6361/201833910}

\bibitem[{{Popov} \& {Postnov}(2013)}]{Popov_Postnov_2013}
{Popov}, S.~B., \& {Postnov}, K.~A. 2013, arXiv e-prints, arXiv:1307.4924, \dodoi{10.48550/arXiv.1307.4924}

\bibitem[{{Press} \& {Schechter}(1974)}]{Schechter1974}
{Press}, W.~H., \& {Schechter}, P. 1974, \apj, 187, 425, \dodoi{10.1086/152650}

\bibitem[{{Qiang} {et~al.}(2022){Qiang}, {Li}, \& {Wei}}]{Qiang+2022}
{Qiang}, D.-C., {Li}, S.-L., \& {Wei}, H. 2022, \jcap, 2022, 040, \dodoi{10.1088/1475-7516/2022/01/040}

\bibitem[{{Ravi} {et~al.}(2015){Ravi}, {Shannon}, \& {Jameson}}]{Ravi_2015}
{Ravi}, V., {Shannon}, R.~M., \& {Jameson}, A. 2015, \apjl, 799, L5, \dodoi{10.1088/2041-8205/799/1/L5}

\bibitem[{{Rodriguez} {et~al.}(2016){Rodriguez}, {Chatterjee}, \& {Rasio}}]{Rodriguez+2016}
{Rodriguez}, C.~L., {Chatterjee}, S., \& {Rasio}, F.~A. 2016, \prd, 93, 084029, \dodoi{10.1103/PhysRevD.93.084029}

\bibitem[{{Ruiter} {et~al.}(2009){Ruiter}, {Belczynski}, \& {Fryer}}]{Ruiter+2009}
{Ruiter}, A.~J., {Belczynski}, K., \& {Fryer}, C. 2009, \apj, 699, 2026, \dodoi{10.1088/0004-637X/699/2/2026}

\bibitem[{{Ryder} {et~al.}(2023){Ryder}, {Bannister}, {Bhandari}, {Deller}, {Ekers}, {Glowacki}, {Gordon}, {Gourdji}, {James}, {Kilpatrick}, {Lu}, {Marnoch}, {Moss}, {Prochaska}, {Qiu}, {Sadler}, {Simha}, {Sammons}, {Scott}, {Tejos}, \& {Shannon}}]{askap_2023}
{Ryder}, S.~D., {Bannister}, K.~W., {Bhandari}, S., {et~al.} 2023, Science, 382, 294, \dodoi{10.1126/science.adf2678}

\bibitem[{{Schwab}(2021)}]{Schwab_2021}
{Schwab}, J. 2021, \apj, 906, 53, \dodoi{10.3847/1538-4357/abc87e}

\bibitem[{{Schwab} {et~al.}(2016){Schwab}, {Quataert}, \& {Kasen}}]{Schwab_2016}
{Schwab}, J., {Quataert}, E., \& {Kasen}, D. 2016, \mnras, 463, 3461, \dodoi{10.1093/mnras/stw2249}

\bibitem[{{Shin} {et~al.}(2023){Shin}, {Masui}, {Bhardwaj}, {Cassanelli}, {Chawla}, {Dobbs}, {Dong}, {Fonseca}, {Gaensler}, {Herrera-Mart{\'\i}n}, {Kaczmarek}, {Kaspi}, {Leung}, {Merryfield}, {Michilli}, {M{\"u}nchmeyer}, {Pearlman}, {Rafiei-Ravandi}, {Smith}, {Stairs}, \& {Tendulkar}}]{Shin_2023}
{Shin}, K., {Masui}, K.~W., {Bhardwaj}, M., {et~al.} 2023, \apj, 944, 105, \dodoi{10.3847/1538-4357/acaf06}

\bibitem[{{Spitler} {et~al.}(2014){Spitler}, {Cordes}, {Hessels}, {Lorimer}, {McLaughlin}, {Chatterjee}, {Crawford}, {Deneva}, {Kaspi}, {Wharton}, {Allen}, {Bogdanov}, {Brazier}, {Camilo}, {Freire}, {Jenet}, {Karako-Argaman}, {Knispel}, {Lazarus}, {Lee}, {van Leeuwen}, {Lynch}, {Ransom}, {Scholz}, {Siemens}, {Stairs}, {Stovall}, {Swiggum}, {Venkataraman}, {Zhu}, {Aulbert}, \& {Fehrmann}}]{Spitler_2014}
{Spitler}, L.~G., {Cordes}, J.~M., {Hessels}, J.~W.~T., {et~al.} 2014, \apj, 790, 101, \dodoi{10.1088/0004-637X/790/2/101}

\bibitem[{{Thompson} \& {Duncan}(1995)}]{Thompson_Duncan_1995}
{Thompson}, C., \& {Duncan}, R.~C. 1995, \mnras, 275, 255, \dodoi{10.1093/mnras/275.2.255}

\bibitem[{{Thornton} {et~al.}(2013){Thornton}, {Stappers}, {Bailes}, {Barsdell}, {Bates}, {Bhat}, {Burgay}, {Burke-Spolaor}, {Champion}, {Coster}, {D'Amico}, {Jameson}, {Johnston}, {Keith}, {Kramer}, {Levin}, {Milia}, {Ng}, {Possenti}, \& {van Straten}}]{Thornton_2013}
{Thornton}, D., {Stappers}, B., {Bailes}, M., {et~al.} 2013, Science, 341, 53, \dodoi{10.1126/science.1236789}

\bibitem[{{Wang} \& {van Leeuwen}(2024)}]{Wang_vanLeeuwen_2024}
{Wang}, Y., \& {van Leeuwen}, J. 2024, \aap, 690, A377, \dodoi{10.1051/0004-6361/202450673}

\bibitem[{{Ye} \& {Fishbach}(2024)}]{Ye_and_Fishbach_2024}
{Ye}, C.~S., \& {Fishbach}, M. 2024, \apj, 967, 62, \dodoi{10.3847/1538-4357/ad3ba8}

\bibitem[{{Ye} {et~al.}(2024){Ye}, {Kremer}, {Ransom}, \& {Rasio}}]{Ye+2024}
{Ye}, C.~S., {Kremer}, K., {Ransom}, S.~M., \& {Rasio}, F.~A. 2024, \apj, 961, 98, \dodoi{10.3847/1538-4357/ad089a}

\bibitem[{{Zhang} {et~al.}(2024){Zhang}, {Dong}, {Rodin}, {Fedorova}, {Huang}, {Li}, {Wang}, {Li}, {Du}, {Xu}, \& {Zhang}}]{Zhang2024}
{Zhang}, K.~J., {Dong}, X.~F., {Rodin}, A.~E., {et~al.} 2024, arXiv e-prints, arXiv:2406.00476, \dodoi{10.48550/arXiv.2406.00476}

\bibitem[{{Zhang} \& {Zhang}(2022)}]{Zhang_Zhang_2022}
{Zhang}, R.~C., \& {Zhang}, B. 2022, \apjl, 924, L14, \dodoi{10.3847/2041-8213/ac46ad}

\end{thebibliography}
\bibliographystyle{aasjournal}

\end{document}